\documentstyle[preprint,aps,tighten]{revtex}

\begin{document}
\draft
\title{Updating the nuclear reaction rate library {\it (REACLIB)}.\\
I. Experimental reaction rates of the proton-proton chain.}
\author{T. E. Liolios\footnote{www.liolios.info}}
\address{Hellenic Military University, Hellenic Army Academy\\
Department of Physical Sciences, Vari 16673, Attica, GREECE }
\maketitle

\begin{abstract}
REACLIB is one of the most comprehensive and popular astrophysical reaction
rate libraries. However, its experimentally obtained rates\ for light
isotopes still rely mainly on the Caughlan \& Fowler (1988) compilation and
have never been updated despite the progress in many relevant nuclear
astrophysics experiments. Moreover, due to fitting errors REACLIB\ is not
reliable at temperatures lower than $10^{7}K\,$.

In this work we establish the formalism for updating the obsolete
Caughlan-Fowler experimental rates of REACLIB. Then we use the NACRE\
compilation and results from the LUNA experiments to update some important
charged-particle induced rates of REACLIB focusing on the proton-proton
chain. The updated rates (available also in digital form)\ can now be used
in the low temperature regime (below $10^{7}K$) which was forbidden to the
old version of REACLIB.
\end{abstract}

\pacs{PACS number(s): 23.60.+e, 26.30.+k, 26.20.+f}

\section{Introduction}

Nuclear reactions have long been established as the engines that provide the
necessary energy for the stars which use that energy to balance the enormous
gravitational pressure of the stellar gas(see Ref.\cite{claytonbook} and
references therein). Every stellar evolution code relies on a nuclear
reaction rate library which is read by the code before any simulation is
performed (see for example Ref.\cite{raumastars,bazan}). The size and
accuracy of that library determines the quality of the relevant simulation
and therefore its nuclear reaction rates should be continually improved and
upgraded .

One of the most comprehensive such libraries is REACLIB, updated\cite
{raustatmod} by the Basel nuclear astrophysics group and used extensively in
small\cite{young} and large\cite{raumastars} scale simulations. According to
its authors, due to fitting errors, REACLIB cannot be safely used at
temperatures lower than $T_{9}=0.01\,$(where $T_{9}$ measured in $%
10^{9}K)\,,\,$despite the fact that nuclear burning of hydrogen isotopes
starts at lower temperatures, such as deuterium burning down the Hayashi
line. That defect of REACLIB is an undesirable consequence of fitting a
single formula to an array of data (see below) which spans many orders of
magnitude (sometimes more than thirty!) with respect to a very extended
temperature range $0.01\leq T_{9}\leq 10.$ In fact even nuclear rates which
are evaluated close to the upper limit of the critical temperature region
are expected to be contaminated with similar (although less severe) fitting
inaccuracies. For example, very important studies where the application of
the old REACLIB rates may be questionable are pre-main and main-sequence
stellar evolution simulations\cite{young,bazan} including the solar
evolution/neutrino studies where temperatures never exceed the value of $%
T_{9}=0.016\,$(the central value of the present sun is roughly $T_{9}=0.0157$%
)$\,$. Therefore it is now obvious that all stellar evolution simulations
which start from a Zero-Age Main Sequence star (ZAMS)\ are forced to apply
REACLIB to temperatures well below $T_{9}=0.01\,$yielding results which may
be inaccurate.

According to the authors of REACLIB most of its charged-particle reaction
rates for light nuclei rely on the compilation of Caughlan and Fowler\cite
{cf88} . However, since the publication of Ref.\cite{cf88} there has been a
very fertile activity in the field of experimental nuclear astrophysics
leading to experiments which for the first time reached deep into the most
effective energy of interaction of astrophysical reactions \cite{lunahe3}.
New reaction rate compilations have appeared either for light nuclei\cite
{nacre} or for heavier ones participating in explosive burning \cite
{iliadis94}. REACLIB has not yet been updated and its light nuclei
experimentally-obtained rates are obsolete. Moreover, the Caughlan-Fowler%
\cite{cf88} rates suffer from another source of inaccuracy since many higher
energy resonances are lumped into one analytical term which is an
undesirable oversimplification for plausible reasons.

The present paper has three objectives:

a) to use the NACRE\ compilation (as well as other more recent experimental
data) in order to partially update the light-isotope experimental reaction
rates of REACLIB. The update is focused on some of the most important
reaction rates of the proton-proton chain which (in their updated version)
can now be used in the critical temperature region $T_{9}<0.01$.

b) to improve the REACLIB fitting accuracy in such a way which would allow
its application to high quality studies related to the destruction of
short-lived nuclei in pre-main sequence stellar evolution.

c) to establish the formalism and techniques which will be used in future
more extended updates of REACLIB.

The layout of the paper is as follows:

In Section II there is a brief description of the formalism used in the
evaluation of light-nuclei thermonuclear reaction rates. In Section III the
main components of REACLIB\ are presented while in Section IV we describe
the methodology adopted in order to accomplish the above mentioned
objectives. In Section V some of the most important reactions of the
proton-proton chain are updated while the results of the present study are
summarized in Section VI.

\section{Calculation of thermonuclear reaction rates}

The thermonuclear reaction rate (TTR) for the binary reaction $X\left(
a,b\right) Y$ is given by the formula

\begin{equation}
r_{aX}=\left( 1+\delta _{aX}\right) ^{-1}N_{a}N_{X}\left\langle \sigma
u\right\rangle  \label{rax}
\end{equation}
where $N_{a},N_{X}$ are the number densities and $\left\langle \sigma
u\right\rangle \,$stands for the reaction rate per pair of particles given
by the formula:

\begin{equation}
\left\langle \sigma u\right\rangle =\sqrt{\frac{8}{\pi \mu }}\frac{1}{\left(
kT\right) ^{3/2}}\int_{0}^{\infty }\sigma \left( E\right) E\exp \left( -%
\frac{E}{kT}\right) dE  \label{su}
\end{equation}
The Kronecker symbol $\delta _{\alpha X}$ takes into account that the
interacting nuclei can be identical.

The cross section $\sigma \left( E\right) ,$which appears in the TRR, can be
non-resonant or resonant according to the range of stellar energies $E$. If
the temperature of the star is such that the integrand goes to zero before
the cross section strikes a resonance the non resonant formalism can be used
by adopting the formula

\begin{equation}
\sigma \left( E\right) =\frac{S\left( E\right) }{E}e^{-2\pi n}  \label{nrcs}
\end{equation}
where $S\left( E\right) \,$is a slowly varying function of energy called the
astrophysical factor and $n$ is the usual Sommerfeld parameter.

On the other hand if the most effective energy of interaction (see Eq.\ref
{moef}) is equal to the energy $\left( E_{r}\right) \,$of a quasi-stationary
state of the ensuing compound nucleus then the cross section exhibits
resonant behavior which can be described by the Breit-Wigner formula:

\begin{equation}
\sigma _{r}\left( E\right) =\frac{\pi }{k^{2}}\omega \frac{\Gamma _{i}\left(
E\right) \Gamma _{f}\left( E\right) }{\left( E-E_{r}\right) ^{2}+\Gamma
\left( E\right) ^{2}/4}  \label{rcs}
\end{equation}
where $\kappa $ is the wave number, $\Gamma _{i}\left( E\right) $ and $%
\Gamma _{f}\left( E\right) $ are the entrance and exit channel partial
widths, respectively, $\Gamma \left( E\right) $ is the total width, and $%
\omega $ is the statistical factor given by

\begin{equation}
\omega =\left( 1+\delta _{12}\right) \frac{\left( 2J+1\right) }{\left(
2J_{1}+1\right) \left( 2J_{2}+1\right) }  \label{omega}
\end{equation}
where $J_{1},J_{2},J,$are the spins of the interacting nuclei and of the
resonance, respectively.

When the Breit-Wigner formula is inserted into Eq. $\left( \ref{su}\right) $
the integrand exhibits maxima at $E_{r}$ and at the most effective energy of
interaction $E_{0}$ given by 
\begin{equation}
E_{0}=0.1220\left( Z_{1}^{2}Z_{2}^{2}A\right) ^{1/3}T_{9}^{2/3}Mev
\label{moef}
\end{equation}
where $Z_{1},Z_{2}$ are the charge numbers of the interacting nuclei and $A$
is the respective reduced mass number $A=A_{1}A_{2}\left( A_{1}+A_{2}\right)
^{-1}$.

The thermonuclear reaction rate (TRR)\ per pair of particles for an isolated
narrow resonance $E_{r}$ is given by\cite{claytonbook} 
\begin{equation}
\left\langle \sigma u\right\rangle _{E_{r}}=\left( \frac{2\pi }{\mu kT}%
\right) ^{3/2}\hbar ^{2}\left( \omega \gamma \right) _{E_{r}}\exp \left( -%
\frac{E_{r}}{kT}\right)  \label{trr}
\end{equation}

where $\left( \omega \gamma \right) _{E_{r}}=\omega \Gamma _{i}\Gamma
_{f}/\Gamma \left( E_{r}\right) $

For light nuclei which capture protons or alpha particles (such as those
studied in the present paper)\ the compound nuclei will be produced at low
excitation energies where the level densities are low. In a such a case the
statistical model (i.e. Hauser Feshbach model) breaks down\cite{iliadis94}
and sometimes overestimates the reaction rates by several orders of
magnitude. Therefore, the total Maxwellian averaged reaction rate $%
N_{A}<\sigma u>$ \thinspace is determined by summing up the contributions of 
$\left( i\right) $ single isolated (narrow) resonances

\begin{equation}
N_{A}<\sigma u>_{r_{i}}=N_{A}\left( \frac{2\pi }{\mu k}\right) ^{3/2}\hbar
^{2}\left( \omega \gamma \right) _{r_{i}}T^{-3/2}\exp \left( -\frac{E_{r_{i}}%
}{kT}\right)  \label{resterm}
\end{equation}
and their single non-resonant (tail) contribution

\begin{equation}
N_{A}<\sigma u>_{nr}=N_{A}\left( \frac{2}{\mu }\right) ^{1/2}\frac{\Delta
E_{0}}{\left( kT\right) ^{3/2}}S_{eff}\exp \left( -\frac{3E_{0}}{kT}\right)
\label{tailterm}
\end{equation}
so that

\begin{equation}
N_{A}<\sigma u>\,=\sum_{i}N_{A}<\sigma u>_{r_{i}}+N_{A}<\sigma u>_{nr}
\label{sumterms}
\end{equation}
where we have used the familiar notation for rates used in many popular
textbooks and articles such as Ref.\cite{claytonbook} and Ref. \cite
{fowler67} and the index $\left( i\right) $ in Eq. $\left( \ref{sumterms}%
\right) $ indicates a particular isolated resonance. Note that $S_{eff}$ is
the effective astrophysical factor which is given as a function of the
experimentally derived zero-energy astrophysical factor and its derivatives $%
\left( S\left( 0\right) ,S^{^{\prime }}\left( 0\right) ...\right) $%
\thinspace by the formula

\begin{equation}
S_{eff}=S\left( 0\right) \left[ 1+\frac{5}{12\tau }+\frac{S^{^{\prime
}}\left( 0\right) }{S\left( 0\right) }\left( E_{0}+\frac{35}{36}kT\right) +%
\frac{1}{2}\frac{S^{^{\prime \prime }}\left( 0\right) }{S\left( 0\right) }%
\left[ E_{0}^{2}+\frac{89}{36}E_{0}kT\right] \right] \,  \label{seff}
\end{equation}
where the $E_{0}$ is the most effective energy of interaction, $\tau $ is
given by $\tau =4.248\left( Z_{1}Z_{2}AT_{9}\right) ^{1/3}$ and $A$ is
reduced mass number: $A=A_{1}A_{2}\left( A_{1}+A_{2}\right) ^{-1}$).

Nuclear astrophysics experiments measure all the components $\left( S\left(
0\right) ,S^{^{\prime }}\left( 0\right) ...\right) $\thinspace of the
effective astrophysical factor $S_{eff}$, the resonance energies $E_{r}\,$%
and the respective partial widths. Those data are then inserted into Eq. $%
\left( \ref{sumterms}\right) $ in order to provide formulas for the
thermonuclear reaction rates which are used in stellar evolution
simulations. However, thermonuclear reaction rate data are more easily used
when they are given in tabular forms\cite{raustatmod,nacre} so that they can
be parameterized into reaction rate libraries by using suitable fitting
formulas \cite{tycho,raustatmod,iliadis94}. Those libraries are then
uploaded directly by the simulation code for stellar evolution and
nucleosynthesis calculations.

\section{Brief description of REACLIB}

REACLIB is a complete library of nuclear reaction rates. Using capital
letters $A,B,C,...$,\thinspace to denote each isotope (parent and daughter
ones) the reaction rate library {\it REACLIB} would consist of the following
components:

DECAYS

Beta decays and electron captures: $A\longrightarrow B$

Photodisintegration and beta delayed neutron emission: $A\longrightarrow B+C$

Inverse triple alpha or beta-delayed two neutron emission: $A\longrightarrow
B+C+D$

BINARY REACTIONS

$A+B\longrightarrow C$

$A+B\longrightarrow C+D$

$A+B\longrightarrow C+D+E$

$A+B\longrightarrow C+D+E+F$

TRIPLE REACTION

$A+B+C\longrightarrow D$

$A+B+C\longrightarrow D+E$

Each rate is described by three lines. The first line indicates: a) the
participating nuclei, b) the source of the reaction, c) the type of reaction
(resonant,non-resonant), d) if the rate is calculated from the inverse
reaction rate or not, e) the Q value of the reaction in MeV. The second and
third line for each rate give the seven fitting coefficients described below.

All reaction rates in REACLIB have been derived by using the
seven-parametric $\left( a_{i},i=1...7\right) $ fitting formula 
\begin{equation}
R_{tot}\left( a_{1}...a_{7};T_{9}\right)
=exp(a_{1}+a_{2}T_{9}^{-1}+a_{3}T_{9}^{-1/3}+a_{4}T^{1/3}+a_{5}T_{9}+a_{6}T_{9}^{5/3}+a_{7}lnT_{9}))
\label{fitform}
\end{equation}
where the reaction rate $R_{tot}\left( a_{1}...a_{7};T_{9}\right) $
corresponds to: $ln2/t_{1/2}$ for decays, $N_{A}<ab>$ for binary reactions, $%
N_{A}^{2}<abc>$ for triple-reactions ($N_{A}$ being Avogadro's number), and $%
T_{9}$ is the temperature in units of GK. According to the above mentioned
formalism $\left( i.e.\,Eq.\left( \ref{resterm}\right) ,Eq.\left( \ref
{tailterm}\right) ,Eq.\left( \ref{sumterms}\right) \right) $ REACLIB splits
the total charged-particle induced rate $R_{tot}\,\,\,$into one non-resonant 
$R_{nr}\left( a_{1}^{nr}...a_{7}^{nr}\right) \,$and $\left( i\right) $ $\,$%
resonant components $R_{r_{i}}\left( a_{1}^{r_{i}}...a_{7}^{r_{i}}\right)
\,\,$denoted respectively by the superscripts $\left( nr\right) $ and $%
\left( r\right) $. Thus the total $R_{tot}$ reaction rate will be

\begin{equation}
R_{tot}=R_{nr}\left( a_{1}^{nr}...a_{7}^{nr}\right) +\sum_{i}R_{r_{i}}\left(
a_{1}^{r_{i}}...a_{7}^{r_{i}}\right)  \label{rtot}
\end{equation}
Note that one doesn't have to include all possible resonances of the
compound nucleus which is formed during charged-particle capture reactions.
It is sufficient to include those isolated (narrow) resonances which are
relevant to the temperature (energy) regime where the rate will be applied
to.

Regarding the use of REACLIB, stellar modelers often apply(\cite{tycho,young}%
, and references therein) REACLIB to the critical region $0.001<T_{9}<0.01\,$%
mentioned in our introduction, by using a very limited reaction network of
light nuclei. Their decision is partly justified by the fact that at such
low temperatures there is only a tiny nuclear energy production, while as
regards nucleosynthesis only very light nuclei are destroyed such as
deuterium, lithium etc. Sometimes nuclear burning at temperatures $%
T_{9}<0.0005\,$is totally disregarded and only decays are considered. It is
obvious that the obsolete Caughlan-Fowler\cite{cf88} rates of REACLIB may
have been a source of errors to all stellar evolution simulations\cite
{raumastars,tycho,young} that have used it.

There are various versions of REACLIB which can be downloaded from Ref.\cite
{tommy}.The most recent version of REACLIB currently available on-line\cite
{tommy} by the Basel group involves the unprecedented number of 5.411
isotopes with a mass numbers range $1\leq A\leq 279.\,$However, its
light-isotope charged-particle experimental rates are still those of Ref.%
\cite{cf88}, which underlines the importance of the present study.

\section{Adopted methodology}

Fitting the REACLIB fitting formula Eq. $\left( \ref{fitform}\right) \,$to
the tabular reaction rate data (e.g. the NACRE\ ones) over the entire
temperature range is not the most accurate approach. The fitting engine is
forced to fit a single formula to an array of data which spans many orders
of magnitude (sometimes more than thirty!) with respect to a very extended
temperature range $0.001\leq T_{9}\leq 10.\,$This approach generates
sometimes a significant error which will be avoided in our study.

On the other hand, there are admittedly more sophisticated mathematical
functions that could fit the data much better than Eq. $\left( \ref{fitform}%
\right) $, such as those given by NACRE\cite{nacre}. However, we must follow
the format of REACLIB otherwise we should be dealing with a different
reaction rate library of a different format (whose fitting function might be
taxing the computer considerably).

We must now turn to the format of the NACRE\ data which must be seriously
taken into consideration. The NACRE\ data are given in the form of an array
of values (rates with respect to temperature plus uncertainties) which is
the result of combining various individual rate components, namely:
non-resonant, narrow resonant+tails, broad resonant and multi--resonant
rates. It is impossible to extract the individual rates from the combined
NACRE\ tabular data although that is necessary in our work due to the format
of REACLIB. Fortunately, the NACRE authors have also derived analytical
approximations to each of these rate components, thus providing a tool for
uploading the new NACRE\ rates into REACLIB. We use the ORIGIN fitting
package which relies on the Levenberg-Marquardt (LM) algorithm (one of the
most powerful and reliable fitting methods) to perform non-linear
regression. Actually, we fit Eq. $\left( \ref{fitform}\right) $ to each of
these analytical approximations only over the temperature range where the
respective rate component plays a non-negligible role to the total rate.
Outside this range the new individual REACLIB rates may not be very reliable
(although the total rate can be safely used). For example, if the
non-resonant (NR)\ rate of a particular reaction is considerably smaller
than the respective (first resonance) R1 rate at temperatures $T>T_{9}^{*}$
then our fitting range for the NR rate for that reaction would be $\left[
0.001,T_{9}^{*}\right] $ provided the relevant NR\ Eq.$\left( \ref{fitform}%
\right) $ behaves asymptotically correctly at $T>T_{9}^{*}\,$(e.g. it is a
decreasing function of temperature)$.$ In such a case we wouldn't recommend
using individually the NR rate of that particular reaction at
temperatures\thinspace $T>T^{*}$. However, at $T<T_{9}^{*}\,\,$that
particular NR\ rate is suitable for all practical applications, where of
course the relevant reaction rate obeys the general rule of Eq. $\left( \ref
{smallrate}\right) $

Normally the fitting procedure would involve fitting Eq. $\left( \ref
{fitform}\right) $ $\,$to the NACRE analytic functions, however we have
decided to fit the exponent of Eq. $\left( \ref{fitform}\right) $ to their
natural logarithms , which is a more accurate approach.

As for the non-resonant rate of resonant reactions, in several cases we have
calculated the non-resonant rates by using the numerical integration
formalism adopted by NACRE\cite{nacre}. This was necessary in order to
verify that the analytic formula given by NACRE has not been misprinted..

Note that in our fit we adopt the assumption made by NACRE\cite{nacre}\ that
all rates $N_{A}<\sigma u>\,$which obey the condition

\begin{equation}
N_{A}<\sigma u>\leq 10^{-25}  \label{smallrate}
\end{equation}
can be considered negligible in practically all astrophysical applications.

We assess the quality and relevance of the updated REACLIB by using the
following three tools:

Firstly, we can ascertain that the new (updated) total REACLIB rates
approximate satisfactorily the total NACRE ones by plotting their relative
difference $RD$ with respect to temperature $\left( T_{9}\right) $:

\begin{equation}
RD\left( T_{9}\right) =100\frac{R_{tot}^{REACLIB\left( new\right) }\left(
T_{9}\right) -R_{tot}^{NACRE}\left( T_{9}\right) }{R_{tot}^{REACLIB\left(
new\right) }\left( T_{9}\right) }\%
\end{equation}

Secondly, we can assess the relevance of updating the REACLIB rates (i.e.
the present work) by plotting the variation of the relative difference
between the old total REACLIB rate and the total NACRE rates, with respect
to temperature $\left( T_{9}\right) $:

\begin{equation}
RD\left( T_{9}\right) =100\frac{R_{tot}^{REACLIB\left( old\right) }\left(
T_{9}\right) -R_{tot}^{NACRE}\left( T_{9}\right) }{R_{tot}^{REACLIB\left(
old\right) }\left( T_{9}\right) }\%
\end{equation}

Thirdly we can assess the deviation between the new and the old REACLIB
rates by plotting their relative difference with respect to time. Provided
that the new REACLIB\ rates approximate well the NACRE\ ones this tool is
also a measure of the relevance of the present update:

\begin{equation}
RD\left( T_{9}\right) =100\frac{R_{tot}^{REACLIB\left( old\right) }\left(
T_{9}\right) -R_{tot}^{REACLIB\left( new\right) }\left( T_{9}\right) }{%
R_{tot}^{REACLIB\left( old\right) }\left( T_{9}\right) }\%
\end{equation}

When necessary, we plot the variation of $RD\,\left( \%\right) $ with
respect to temperature for two different temperature regimes: The first
regime is relevant to solar evolution simulations while the second one
covers the entire temperature range given by NACRE.

In each figure caption we also include the accuracy $\left( n\%\right) $ of
the analytical approximation given by NACRE\thinspace (see Appendix B of Ref.%
\cite{nacre}). Therefore all the updated REACLIB reaction rates relying on
the NACRE compilation carry an inherent fitting error of $\left( n\%\right)
. $

\section{Updating the reactions of the proton-proton chain}

\subsection{$^{1}H\left( p,\nu e^{+}\right) \,^{2}H$}

This reaction is of paramount importance to stellar evolution (and
especially to solar evolution/neutrino) studies therefore its thermonuclear
reaction rate at relevant temperatures should be as accurate as possible.

Adopting the same formula for the astrophysical factor as the one given by
NACRE\cite{nacre}:

\[
S\left( E\right) =3.94\times 10^{-25}\times \left( 1+11.7E+75E^{2}\right)
\,MeV\,b 
\]
we numerically integrate Eq. $\left( \ref{su}\right) \,$and then fit Eq. $%
\left( \ref{fitform}\right) \,$to the derived tabular data over the region $%
0.001<T_{9}<0.1$. We confined our fit to a much shorter range $%
0.001<T_{9}<0.02\,$but the accuracy didn't improve. It is obvious from Fig.1
that at temperatures $T_{9}<0.1$ the old REACLIB\ values are very close to
the new ones, therefore the old $^{1}H\left( p,\nu e^{+}\right) ^{2}H$ rate
need not be updated as regards the solar evolution zone. However, although
at temperatures $T_{9}<0.1$ both the old and the new REACLIB rates
approximate well the NACRE data, at higher temperatures the old REACLIB rate
significantly deviates from the NACRE\ one. This deviation is particularly
relevant to explosive hydrogen burning simulations which sometimes are
performed using the REACLIB library (e.g. the TYCHO\cite{tycho} code, which
is based on the REACLIB\ library, is equipped with explosive burning
simulation engines). Therefore we recommend the use of the present updated
reaction rate parameters over the entire spectrum of temperatures: $%
0.001<T_{9}<10$.

{\bf Figure 1}. $^{1}H\left( p,\nu e^{+}\right) \,^{2}H:\,$The variation
RD(\%)\ of the relative difference between the REACLIB rates and the values
obtained by numerical integration of Eq. $\left( \ref{su}\right) \,$using
the NACRE\ data. The solid curve represents the RD\ between the new REACLIB\
rate and the NACRE\ one while the dotted one represents the RD between the
old REACLIB rate and the NACRE one.\thinspace $\left( n=3\%\right) $

\subsection{$^{2}H\left( p,\gamma \right) \,^{3}He$}

NACRE\ specifies two temperature regimes and fits two different functions
for the reaction rates. However, REACLIB cannot follow the same format.
Instead, we fitted the REACLIB rate formula to the NACRE tabulated rates and
we found that the NACRE rates and those given by the new REACLIB are almost
identical for the entire low-temperature regime. However, we observe a 37\%
maximum relative difference between the old REACLIB rates and the NACRE\
ones while at the solar temperature regime, in particular, the RD is roughly
25\%. All deuterium burning studies which have relied on the old REACLIB
should take that observation seriously into account.

{\bf Figure 2}. $^{2}H\left( p,\gamma \right) \,^{3}He:$ The variation of
the relative difference RD between the old REACLIB\ rates and the NACRE\
ones with respect to temperature $\left( n=3\%\right) $

{\bf Figure 3.} $^{2}H\left( p,\gamma \right) \,^{3}He:\,$The variation of
the relative difference RD between the new REACLIB\ rates and the NACRE\
ones with respect to temperature $\left( n=3\%\right) $

\subsection{$^{2}H\left( d,\gamma \right) \,^{4}He$}

Fitting the REACLIB formula (i.e. Eq. $\left( \ref{fitform}\right) $) to the
tabulated data of the NACRE compilation yields very satisfactory results.
The new REACLIB formula fits excellently the NACRE tabular data and
therefore plotting the variation of the relevant RD with respect to
temperature is unnecessary.. Instead, we plot the variation of the relative
difference between the old and the new REACLIB\ rates with respect to
temperature. Fig. 4 shows that the old REACLIB rate in the solar regime can
be up to 15\% larger than the rate predicted by NACRE while this discrepancy
is fixed by the new fit. The updated REACLIB\ rate is practically the same
as the NACRE one in the same regime. However, as shown in the same figure at
larger temperatures the new REACLIB rate (i.e. the NACRE rates) are
significantly larger than the old ones.

{\bf Figure 4.}$^{2}H\left( d,\gamma \right) \,^{4}He:$The variation of the
relative difference RD between the new and old REACLIB\ rates with respect
to temperature $\left( n=3\%\right) $

\subsection{$^{2}H\left( d,n\right) \,^{3}He$}

The new REACLIB formula fits excellently the NACRE tabular data and thus, as
in the case of $^{2}H\left( d,\gamma \right) \,^{4}He$ reaction we only plot
the variation of the relative difference between the old and the new
REACLIB\ rates with respect to temperature.

According to Fig. 5 in the solar regime the old REACLIB rates are up to 12\%
smaller than the new (or NACRE) ones while at higher temperatures this
effect is reversed and the old rates become larger than the new ones (up to $%
90\%$).

{\bf Figure 5.}$^{2}H\left( d,n\right) \,^{3}He:$The variation of the
relative difference RD between the new and old REACLIB\ rates with respect
to temperature $\left( n=4\%\right) $

\subsection{$^{2}H\left( d,p\right) \,^{3}H$}

The new REACLIB formula fits excellently the NACRE tabular data. According
to Fig. 6 in the solar regime the old REACLIB rates are up to 10\% smaller
than the new (or NACRE) ones while at higher temperatures this effect is
reversed and the old rates become larger than the new ones (up to $90\%$)

{\bf Figure 6.} $^{2}H\left( d,p\right) \,^{3}H$: $\,$The variation of the
relative difference RD between the new and the old REACLIB\ rates with
respect to temperature.$\left( n=5\%\right) $

\subsection{$^{3}He\left( ^{3}He,2p\right) \,^{4}He$}

For this reaction we don't rely on NACRE data in order to produce its
REACLIB rate. The LUNA collaboration has managed to lower the beam energy of
their experiment so much that they have recently evaluated the relevant
astrophysical factor with the highest precision ever. Therefore we follow
the most reliable procedure of numerically integrating the thermonuclear
reaction rate integral, a method followed by NACRE as well.

Then we fit Eq. $\left( \ref{fitform}\right) \,$to the array of numerical
data. The new REACLIB formula represents the data very satisfactorily and
according to Fig. 7 there is a notable deviation from the old REACLIB
formula, which may have a non-negligible effect on solar evolution
simulations using REACLIB.

In Fig.7 we plot the variation of the RD\ between the REACLIB\ rates (old
and new)\ and the rate obtained by numerically integrating Eq. $\left( \ref
{su}\right) \,$using the most recent LUNA data. We observe that the RD
between the new REACLIB and the LUNA rates (solid curve) is consistently
smaller than the respective RD (dotted curve) between the old REACLIB and
the LUNA rates. Especially in the solar regime the old REACLIB rate deviates
from the LUNA one by up to 7\% whereas, in the same region, the RD between
the new REACLIB rate and the LUNA one is less than 1\%.

{\bf Figure 7.} $^{3}He\left( ^{3}He,2p\right) \,^{4}He$ The variation of
the relative difference RD between the (old/new) REACLIB\ rates and that
obtained by using the LUNA\ data with respect to temperature. The solid
(dotted) curve represents the RD between the new (old) REACLIB\ rate and the
LUNA one.

\subsection{$^{3}He\left( \alpha ,\gamma \right) \,^{7}Be$}

The new REACLIB formula fits excellently the NACRE tabular data. According
to Fig.8 in the solar region the new REACLIB\ rate approximates the NACRE\
rate with an accuracy of 1\% or better, while the old REACLIB rate (see Fig.
9) can be up to 2.5\% larger than the NACRE one. According to Fig. 10, which
shows the deviation between the old and the new REACLIB\ rates, the new
REACLIB\ rate is approximately the same as the old REACLIB\ one. However,
due to the importance of the $^{3}He\left( \alpha ,\gamma \right) \,^{7}Be$
reaction in the solar neutrino studies we recommend using the new updated
REACLIB\ rate.

{\bf Figure 8.}$\,^{3}He\left( \alpha ,\gamma \right) \,^{7}Be:$ The
variation of the RD between the new REACLIB\ rate and the NACRE\ one with
respect to temperature. $\left( n=6\%\right) $

{\bf Figure 9.}\thinspace $^{3}He\left( \alpha ,\gamma \right) \,^{7}Be$:
The variation of the RD between the old REACLIB\ rate and the NACRE\ one
with respect to temperature. $\left( n=6\%\right) $

{\bf Figure 10.}$\,^{3}He\left( \alpha ,\gamma \right) \,^{7}Be:$The
variation of the RD between the old and the new REACLIB\ rates with respect
to temperature. $\left( n=6\%\right) $

\subsection{$^{6}Li\left( p,\gamma \right) \,^{7}Be$}

The new REACLIB rate approximates much better the NACRE\ rate than the old
one. According to Figs. 11 and 12 in the solar region the old REACLIB rate
can differ from the NACRE\ rate by up to 80\% while the respective
discrepancy for the new REACLIB is always less than 2\%.

{\bf Figure 11.} $^{6}Li\left( p,\gamma \right) \,^{7}Be:$The variation of
the RD between the old REACLIB\ rate and the NACRE\ one with respect to
temperature. $\left( n=7\%\right) $

{\bf Figure 12.}$\,^{6}Li\left( p,\gamma \right) \,^{7}Be:$The variation of
the RD between the new REACLIB\ rate and the NACRE\ one with respect to
temperature. $\left( n=7\%\right) $

\subsection{$^{6}Li\left( p,\alpha \right) \,^{3}He$}

REACLIB distinguishes two components for that rate: a non-resonant and a
resonant one while NACRE\ adopts a single non-resonant fit. By fitting Eq. $%
\left( \ref{fitform}\right) $ to the single analytic formula given by NACRE
we observe a very satisfactory representation of all the NACRE\ tabulated
data. In Fig.13 we compare the new REACLIB\ fit and the old two-component
one where a minor deviation between the new fit and old one is observed.
Accordingly we recommend a single non-resonant REACLIB\ formula for the
updated library

{\bf Figure 13.}$\,^{6}Li\left( p,\alpha \right) \,^{3}He:$The variation of
the RD between the old REACLIB\ rate and the new one with respect to
temperature. $\left( 2\%\right) $

\subsection{$^{7}Li\left( p,\gamma \right) \,^{8}Be$}

This reaction is missing from REACLIB and so is the relevant ensuing decay $%
^{8}Be\rightarrow \,^{4}He+\,^{4}He,\,$ therefore we cannot compare the new
REACLIB rates to the old ones. The importance of this reaction to the PPII\
chain is that it is in competition with the $^{7}Li\left( p,\alpha \right)
\,^{4}He$ reaction, the latter being much more important to the solar
evolution studies, of course. The $^{8}Be$ produced in the $^{7}Li\left(
p,\gamma \right) \,^{8}Be$ reaction, which is unstable and decays into two
alpha particles in $2.6\times 10^{-16}s,$ is extremely important to the
triple alpha reaction as well. Due to the importance of that reaction we
will defer its study (and/or update) to a later paper where we will also
investigate the effects of its absence on the simulations that have used
REACLIB.

\subsection{$^{7}Li\left( p,\alpha \right) \,^{4}He$}

According to Fig.14 the non-resonant rate dominates at temperatures $T_{9}<4$%
. We have compared the NACRE\ rates and those given by the old REACLIB and
have found that their small differences are within the relevant
uncertainties. Therefore, no update was deemed necessary for that reaction.

{\bf Figure 14.}$\,^{7}Li\left( p,\alpha \right) \,^{4}He:$ The logarithms
of the NACRE\ rates (resonant and non-resonant) with respect to temperature. 
$\left( 6\%\right) $

\subsection{$^{7}Li\left( \alpha ,\gamma \right) \,^{11}B$}

NACRE\ evaluates the rates using a non-resonant (NR), a resonant (R1) and a
multi-resonant (MR) rate (see Fig.15) while REACLIB relies only on a NR\ and
a R1 rate. In Fig.16 we plot the variation of RD between the old/new
REACLIB\ rates and the NACRE one for all relevant temperatures. We do not
include an inset figure for the solar regime as that particular reaction is
irrelevant to solar evolution studies. It is obvious that the updated rates
approximate the NACRE\ ones better than the old REACLIB ones.

{\bf Figure 15.} $^{7}Li\left( \alpha ,\gamma \right) \,^{11}B$ : The
logarithms of the NACRE\ rates (resonant, non-resonant and multiresonant)
with respect to temperature.$\left( n=17\%\right) $

{\bf Figure 16.}\thinspace $^{7}Li\left( \alpha ,\gamma \right) \,^{11}B$:
The variation of the RD between the old (solid curve) / new(dotted curve)
REACLIB\ rates and the NACRE one with respect to temperature.$\left(
n=17\%\right) $

\subsection{$^{7}Be\left( p,\gamma \right) \,^{8}B$}

The NACRE non-resonant data for this reaction have been superseded by more
recent ones\cite{schumann}. According to NACRE\cite{nacre} the recommended
S-factor at zero energy is $S_{17}\left( 0\right) =21\pm 2\,eV\,b\,$while
according to Ref. \cite{schumann} it is should be $S_{17}\left( 0\right)
=18.6\pm 1.2\,eV\,b.\,$Despite the notable difference in the zero-energy
astrophysical factor we decided to use the NACRE\ data for consistency.
However, it should be noted that for high quality solar neutrino
calculations the more recent value\cite{schumann}\thinspace \thinspace
should be adopted which would lead to an $11.4\%$ decrease in the relevant
non-resonant rate.

The new REACLIB fitting approximates the NACRE tabular data better than the
old one in the range $0.002<T_{9}<2.2$ while at higher temperatures the old
REACLIB rate constitutes a better approximation. Due to the large
uncertainties involved at such high temperatures we recommend the use of the
updated REACLIB\ rates over the entire spectrum of temperatures. In Fig.17 $%
\,$we observe that the NR\ component of the rate dominates the R1 one
everywhere while in Fig.18 we plot: a) the variation of the RD between the
new REACLIB rate and the NACRE one (solid curve) and b): the variation of
the RD between the old and the new REACLIB rates (dotted curve). It is
obvious that the new REACLIB\ rates are generally more reliable than the old
ones especially at the solar evolution regime.

{\bf Figure 17.} $^{7}Be\left( p,\gamma \right) \,^{8}B:$The logarithms of
the NACRE\ rates (resonant, non-resonant and multiresonant) with respect to
temperature. $\left( n=3\%\right) $

{\bf Figure 18.} $^{7}Be\left( p,\gamma \right) \,^{8}B:$The variation of
the RD between the new REACLIB rate and the NACRE one (solid curve) and the
variation of the RD between the old and the new REACLIB rates (dotted
curve). $\left( n=3\%\right) $

\section{CONCLUSIONS}

The nuclear reaction rate library REACLIB is one of the most comprehensive
and popular ones and is extensively used in stellar evolution and
nucleosynthesis simulations. In the present study some very important
light-isotope charged-particle experimental rates of REACLIB have been
updated using the NACRE$\cite{nacre}\,$compilation and results from the LUNA%
\cite{lunahe3} experiments. We have focused on the most important reaction
rates of the proton-proton chain while the updated REACLIB rates can be used
at temperatures which were forbidden to the old ones. The deviation between
the new and the old REACLIB rates is sometimes significant especially at the
low temperature regime $\left( 0.001<T_{9}<0.01\right) $ of deuterium
burning $^{2}H\left( d,\gamma \right) \,^{4}He,^{2}H\left( d,p\right)
\,^{3}H,^{2}H\left( d,n\right) \,^{3}He\,\,$(where the old REACLIB\ rates
were unreliable).\thinspace Another notable deviation is that the most
important reaction $^{1}H\left( p,\nu e^{+}\right) \,^{2}H\,\,$appears to be
faster in the updated REACLIB than in the old one. The effects of these
deviations on explosive hydrogen burning and big-bang nucleosynthesis should
be carefully investigated by adopting successively the old and the new
REACLIB libraries in relevant simulations (currently under study by the
author).

Another improvement of the new REACLIB\ rates (which are also available in
the same digital form like the old ones) is that we have improved their
fitting accuracy in such a way which allows their application to pre-main
sequence stellar evolution. Finally we have established the formalism and
techniques which will be used in future more extended updates of REACLIB
(soon to appear by the author).

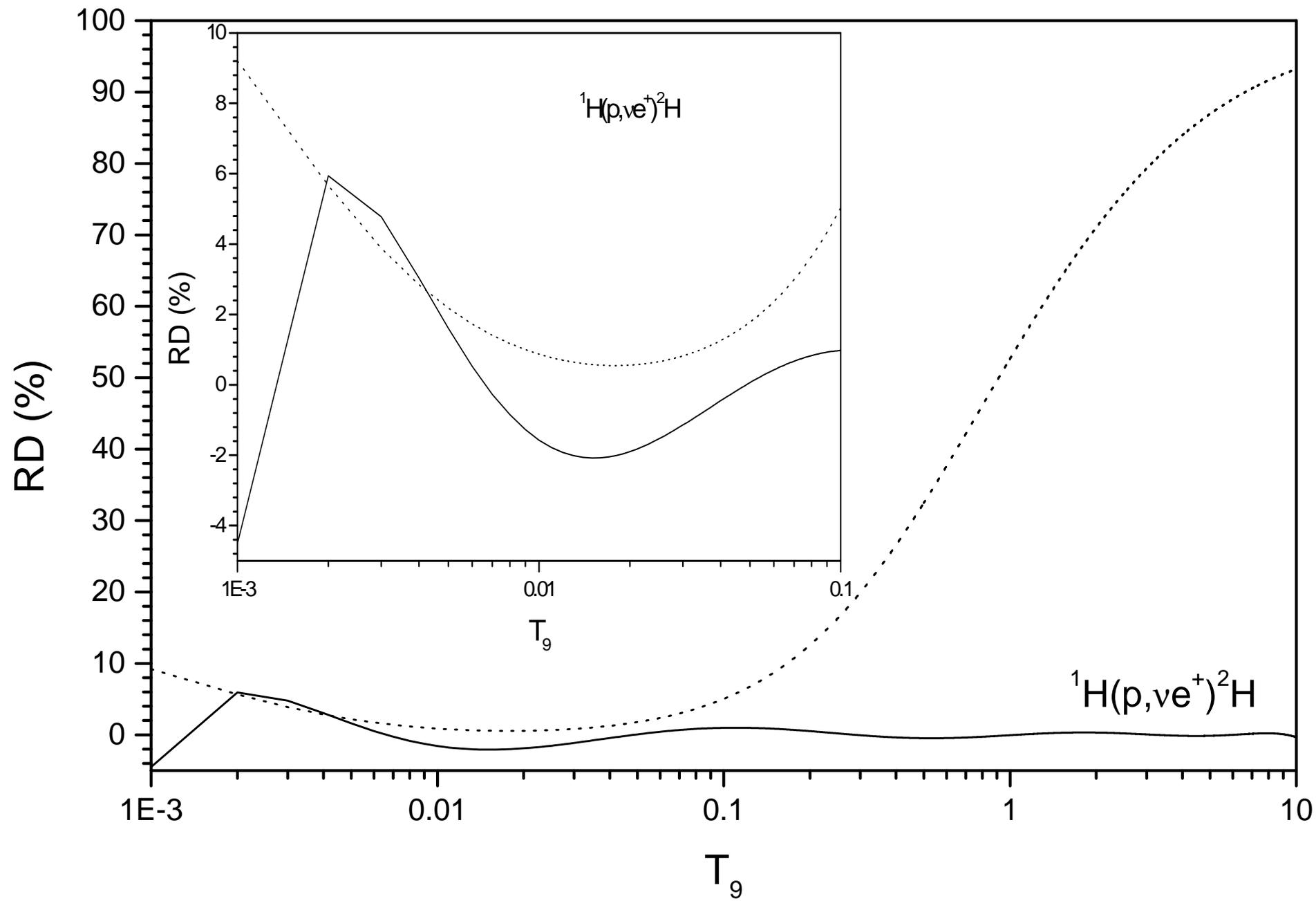
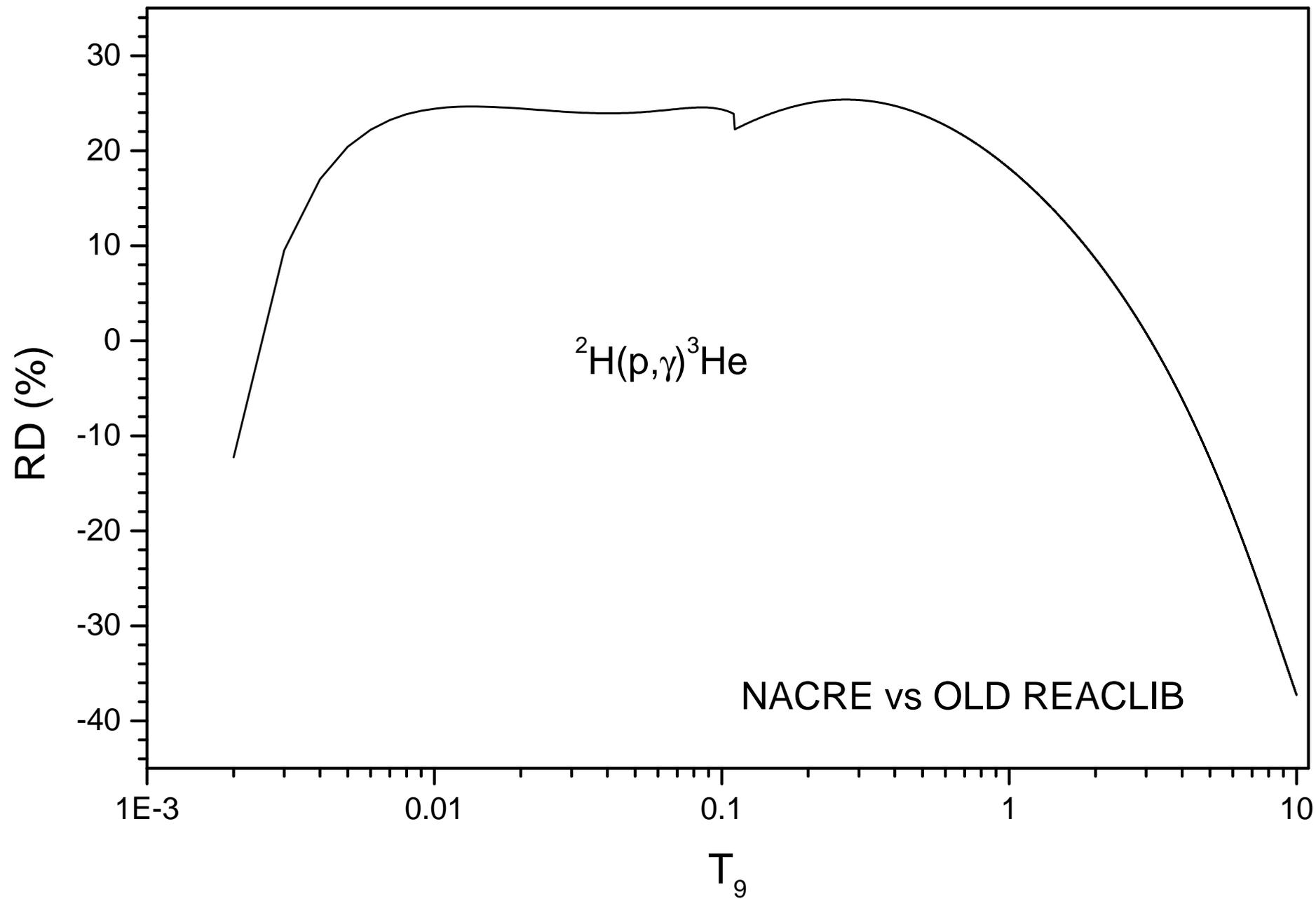
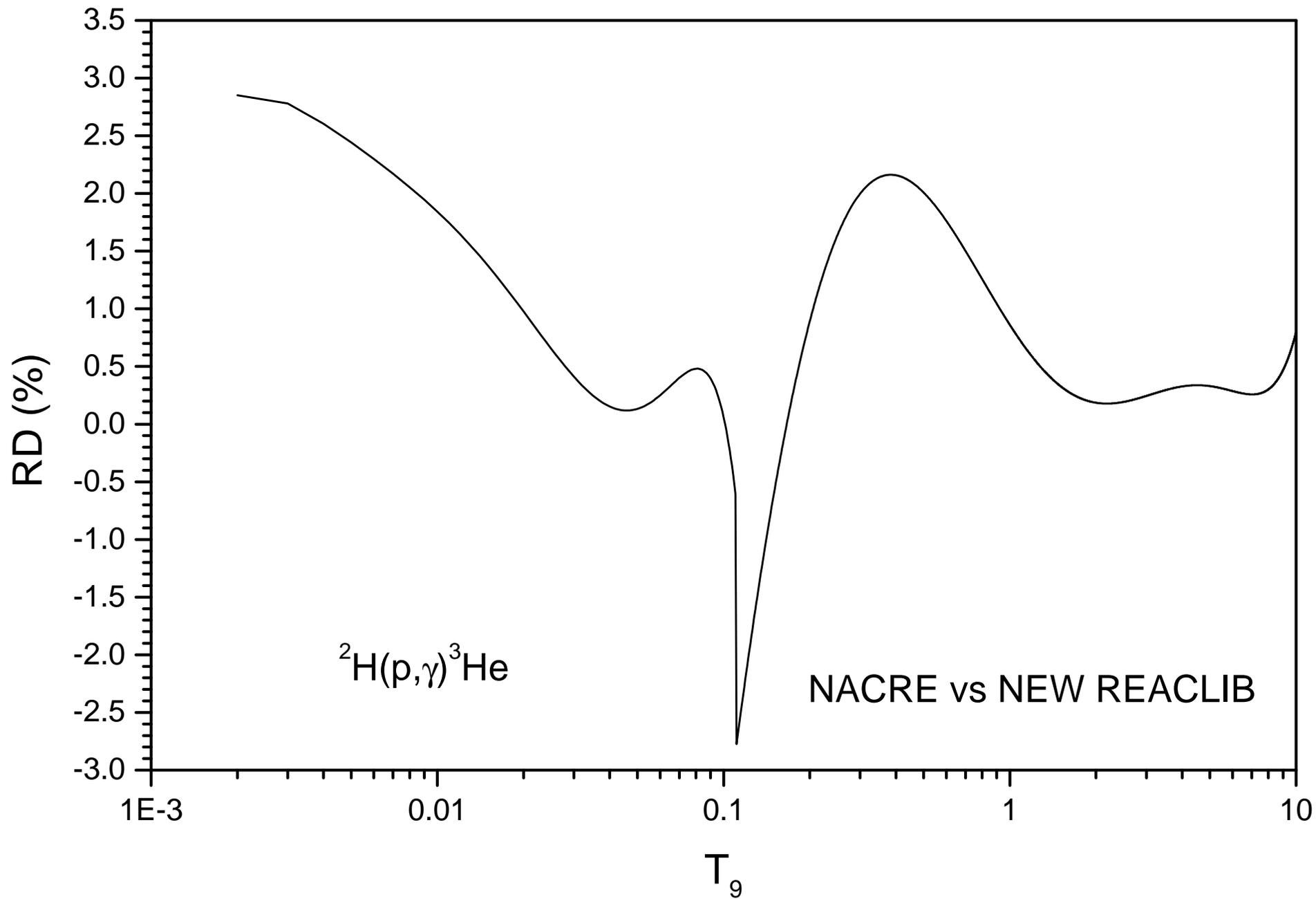
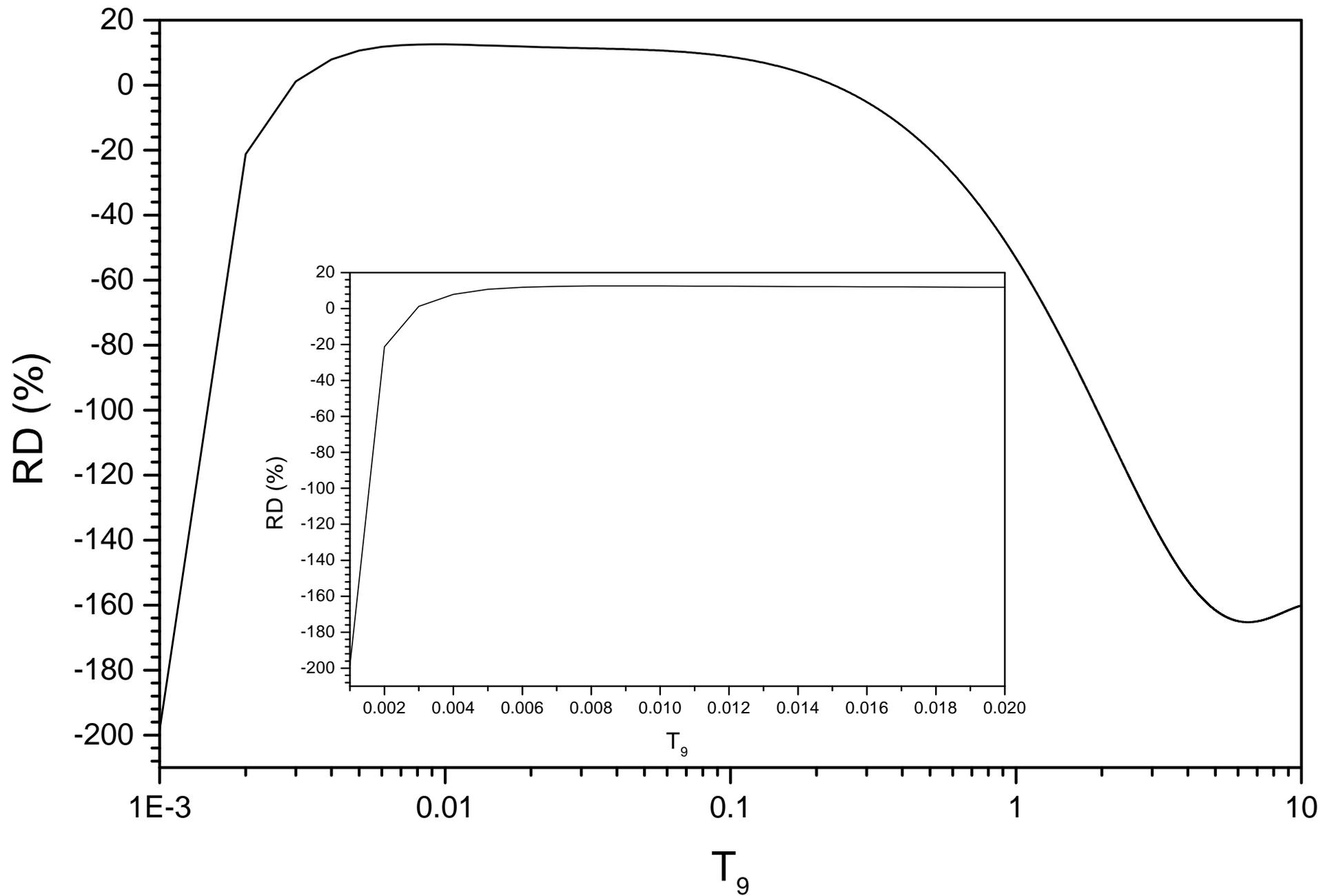
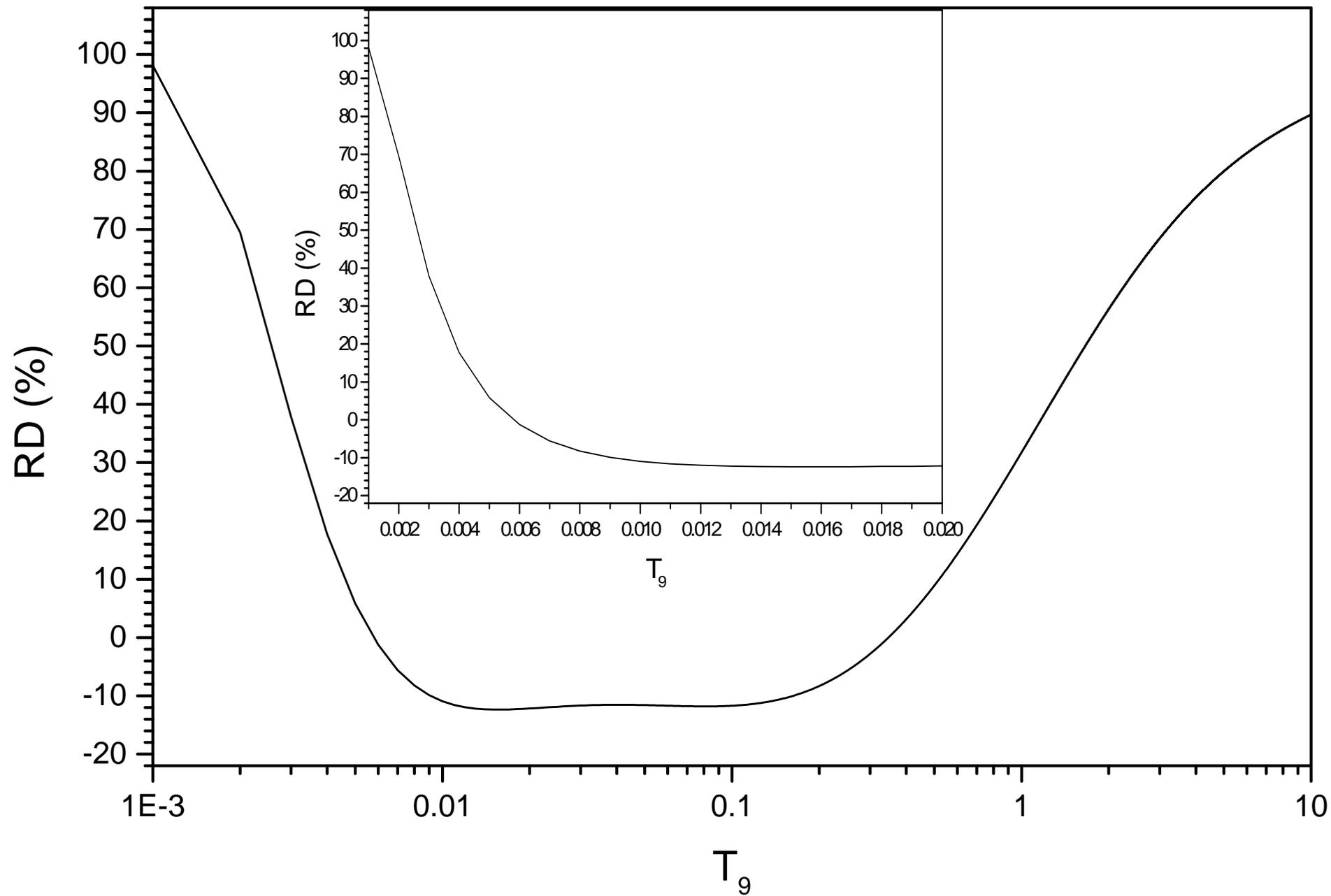
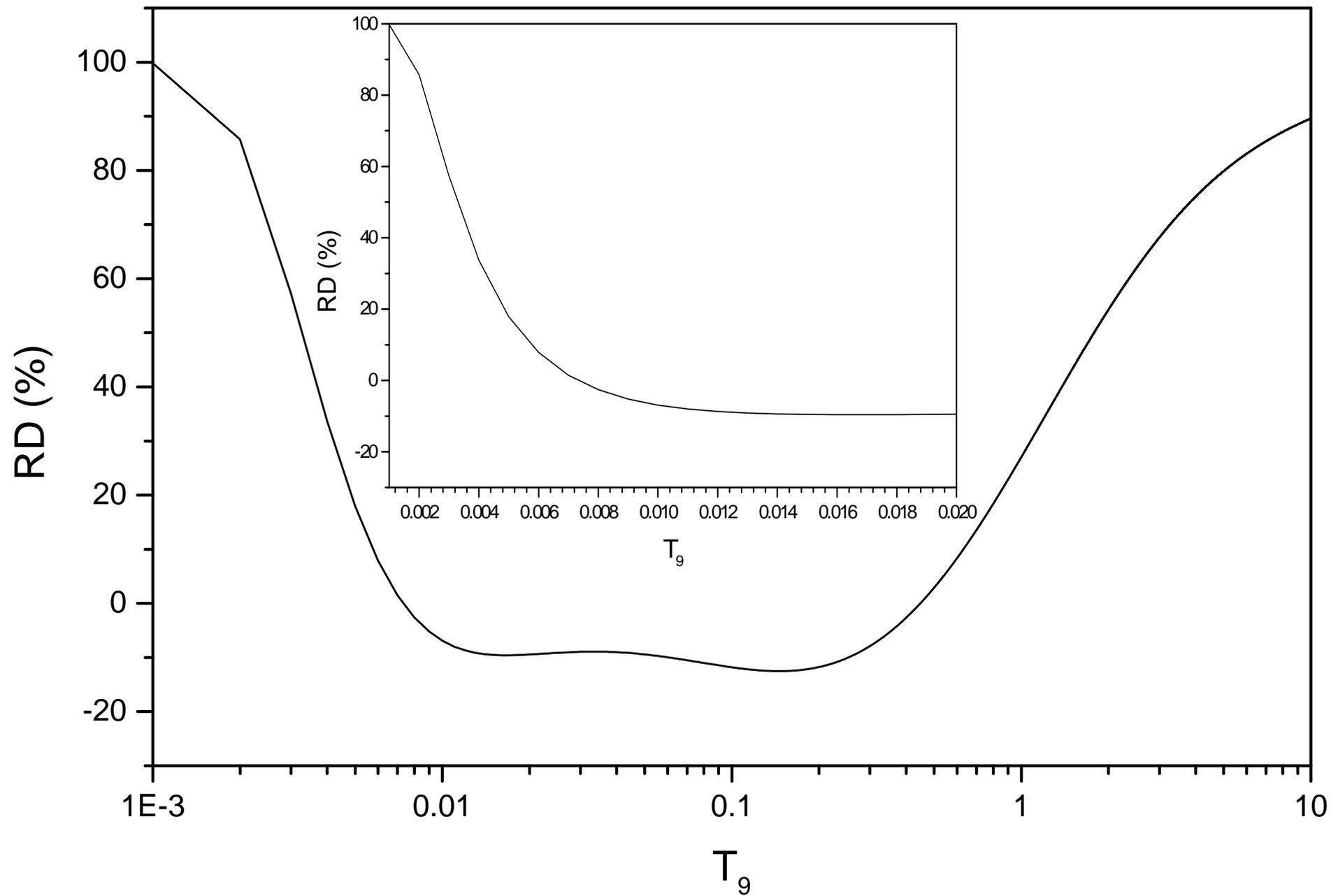
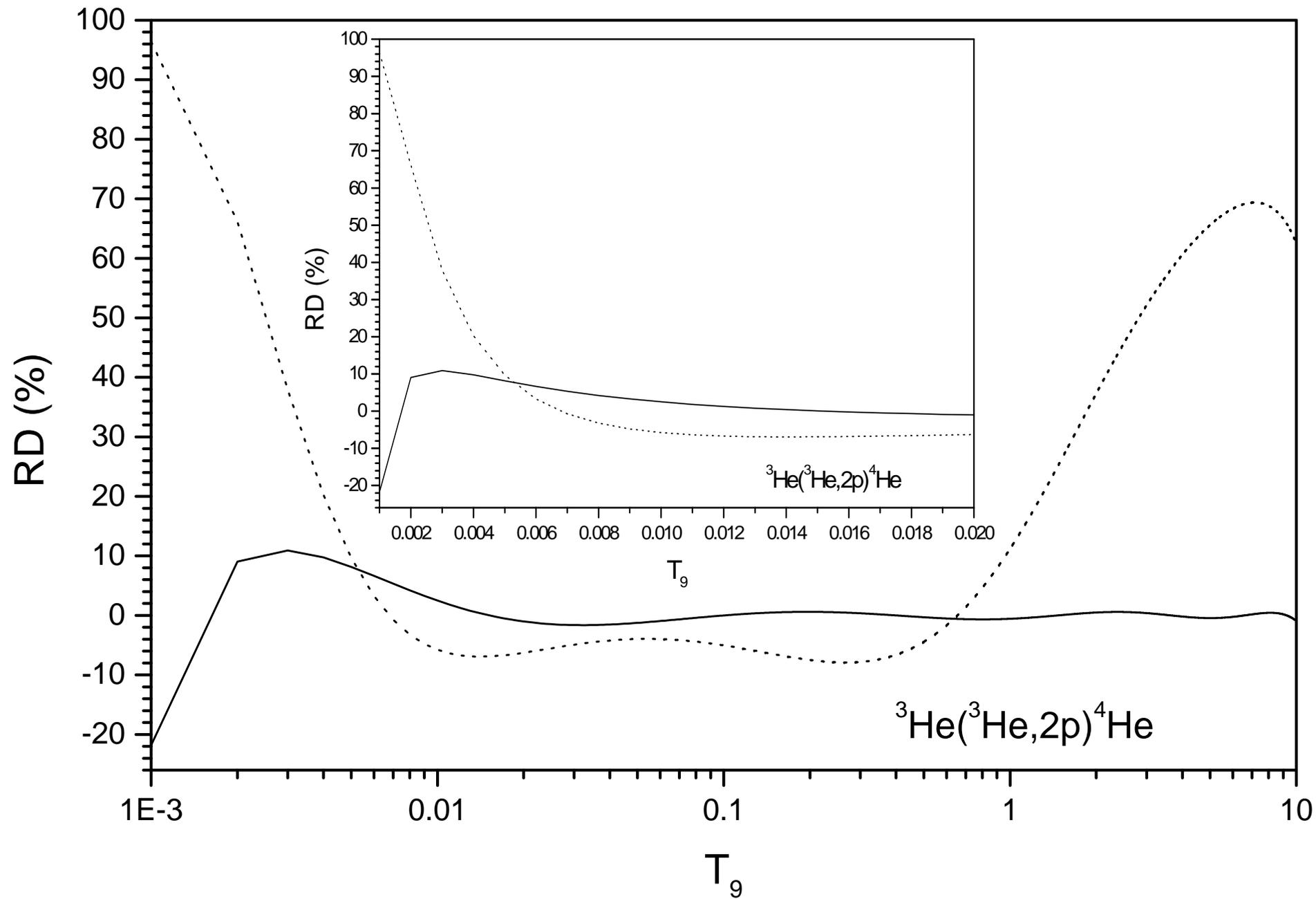
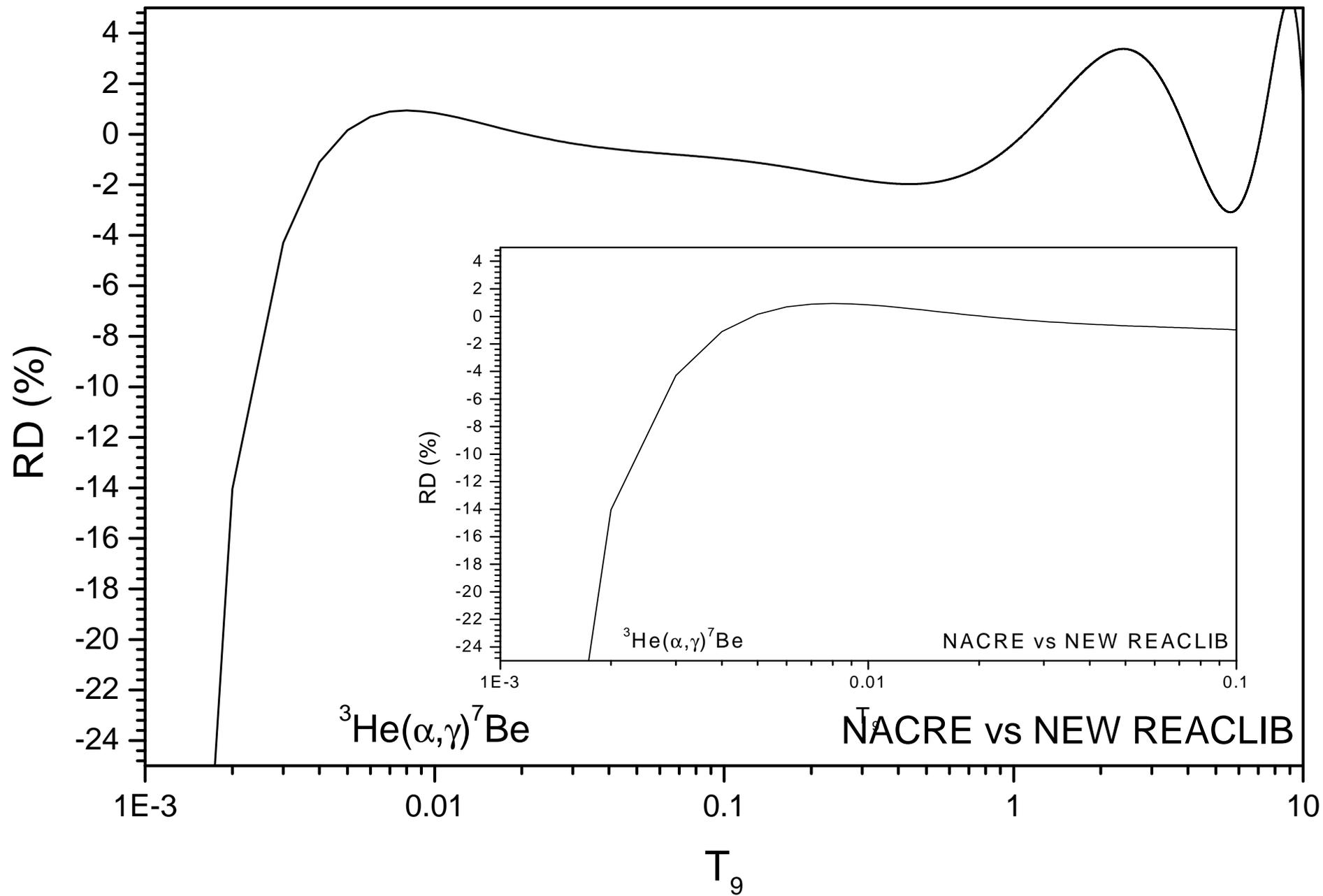
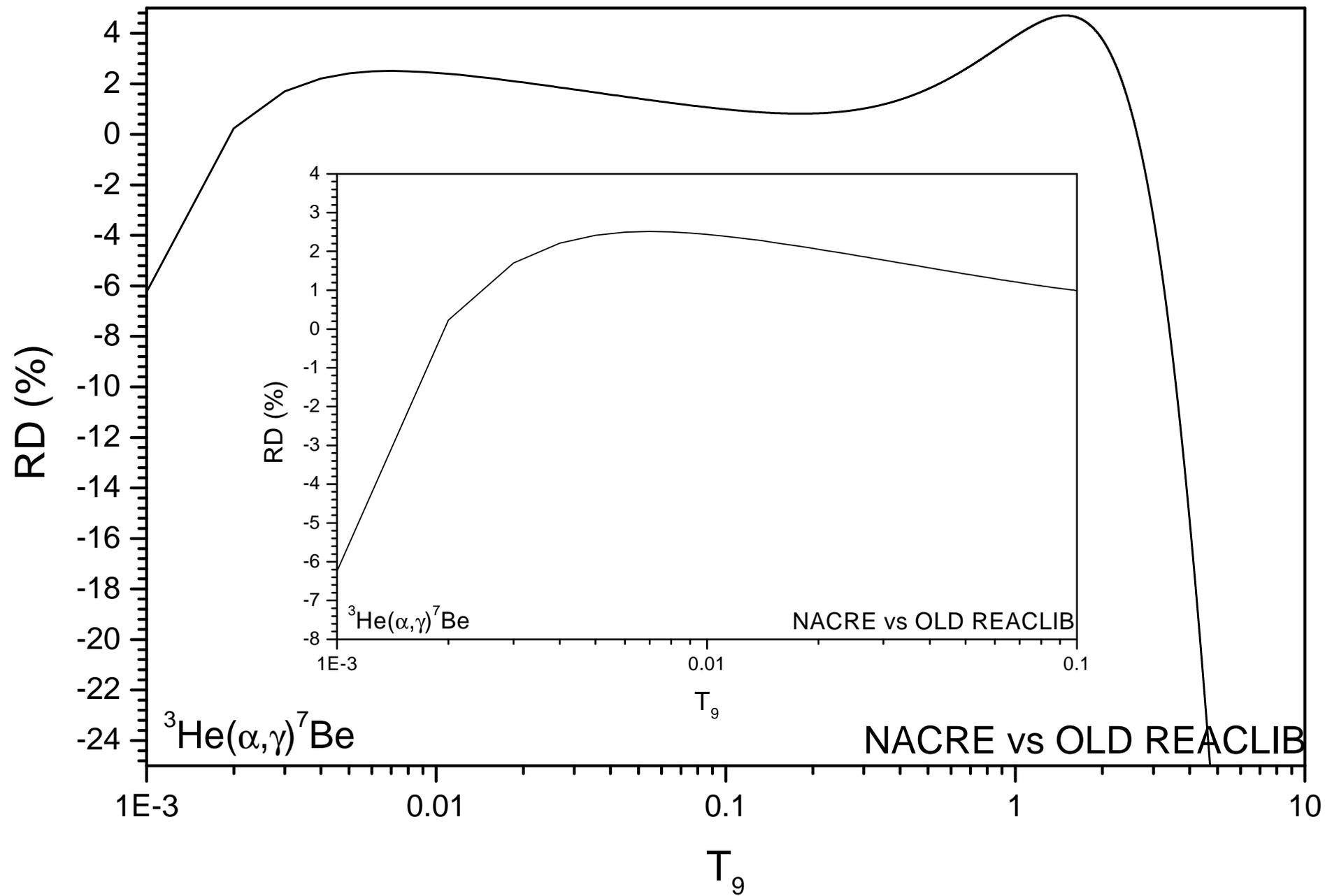
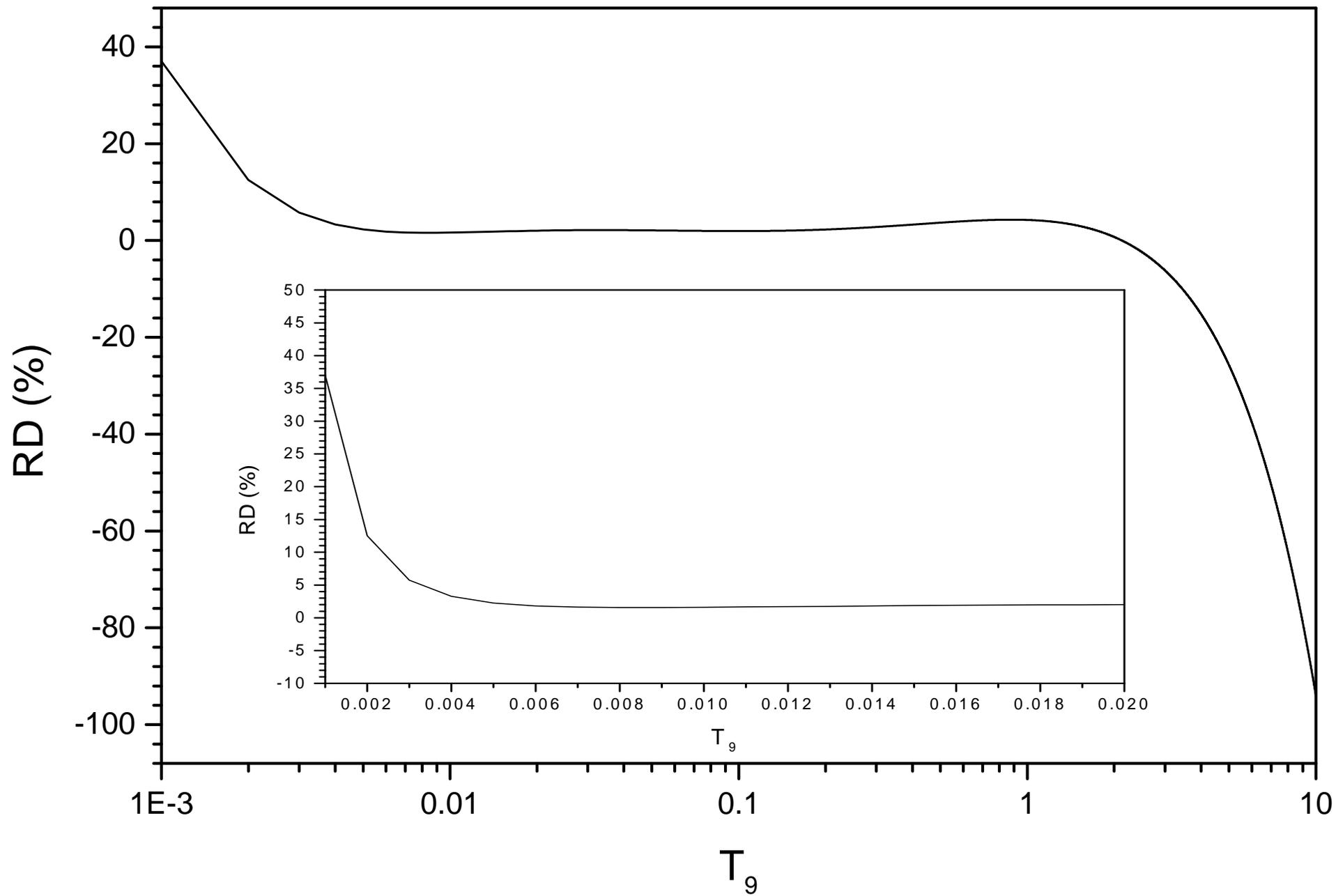
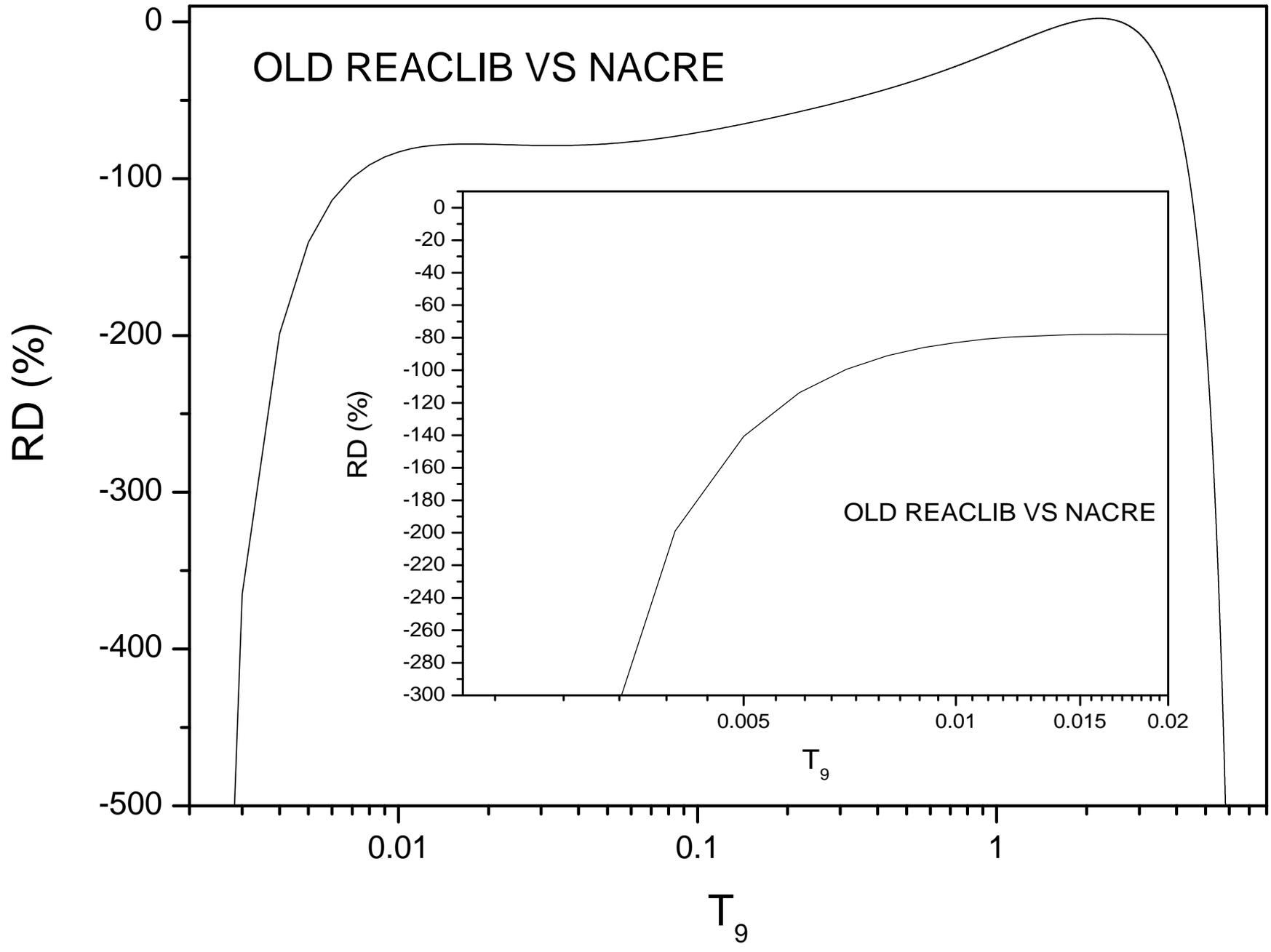
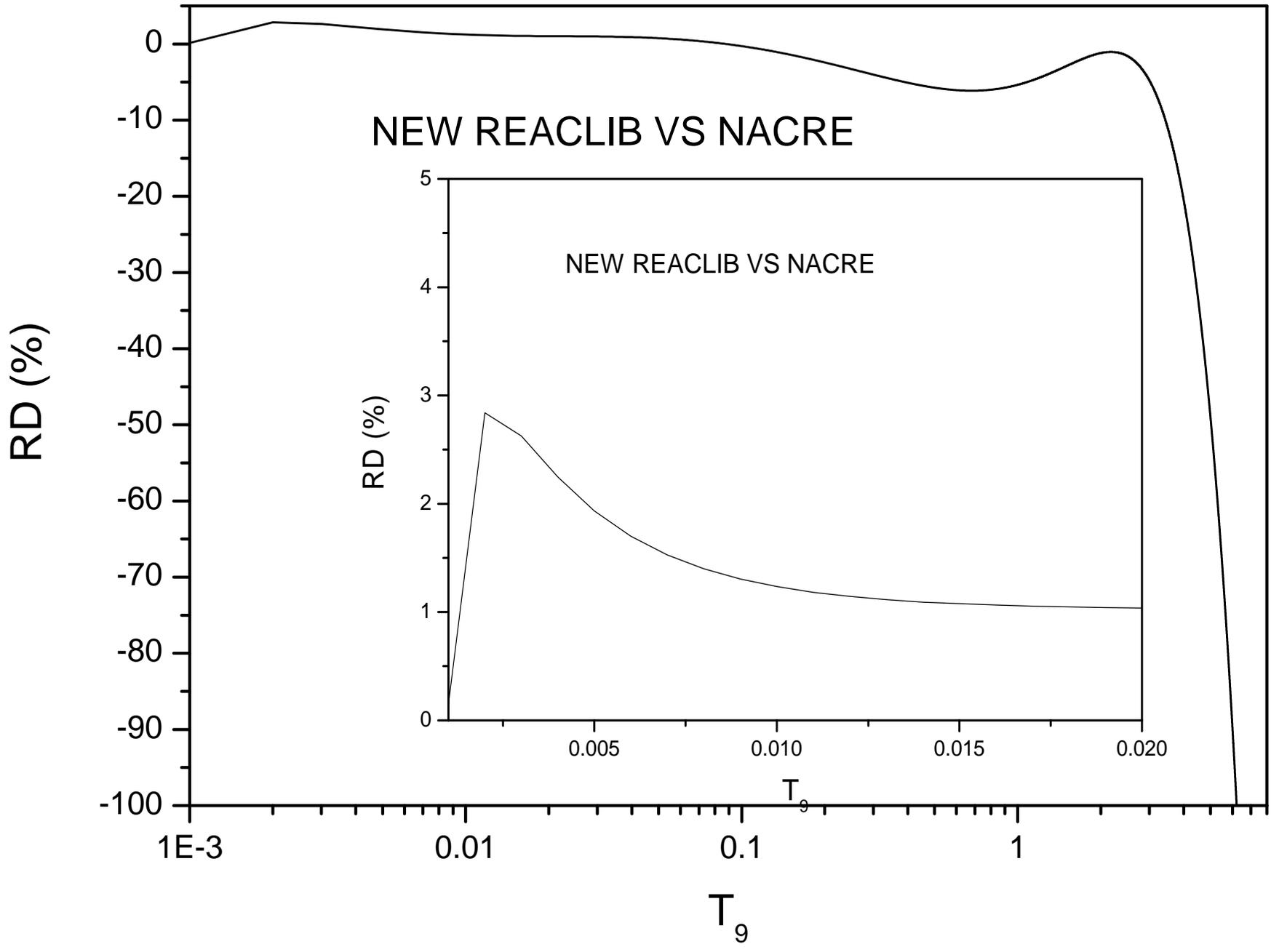
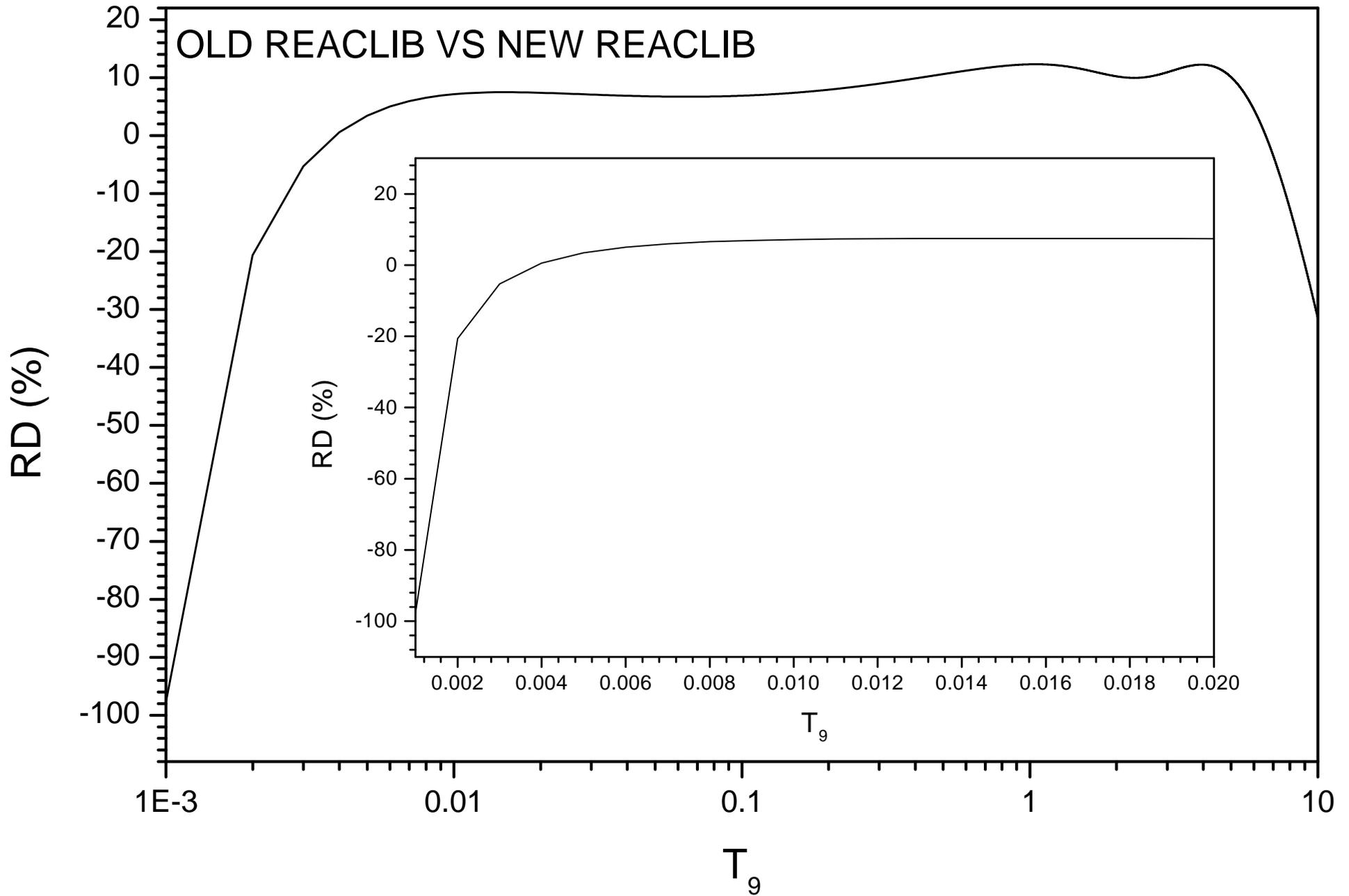
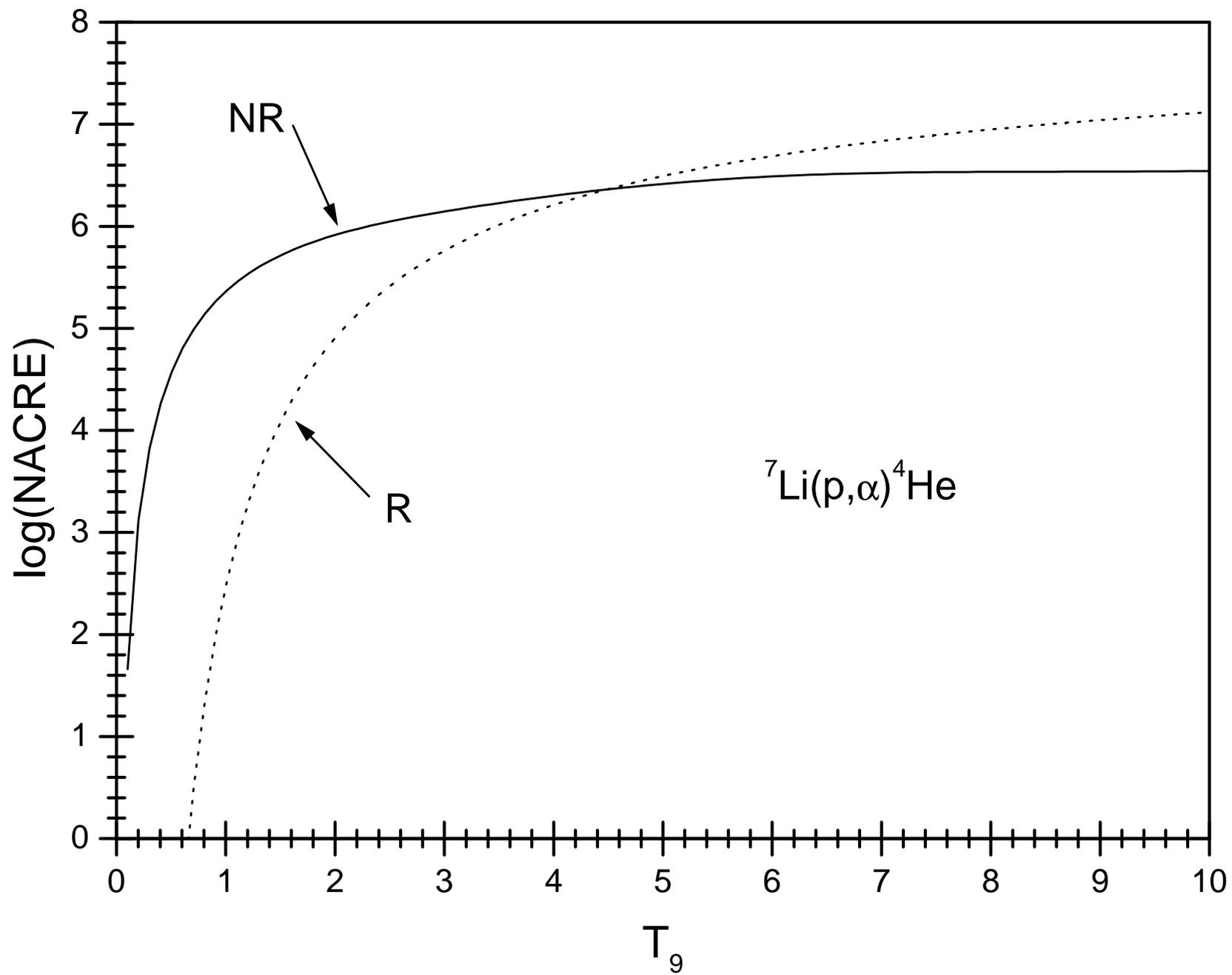
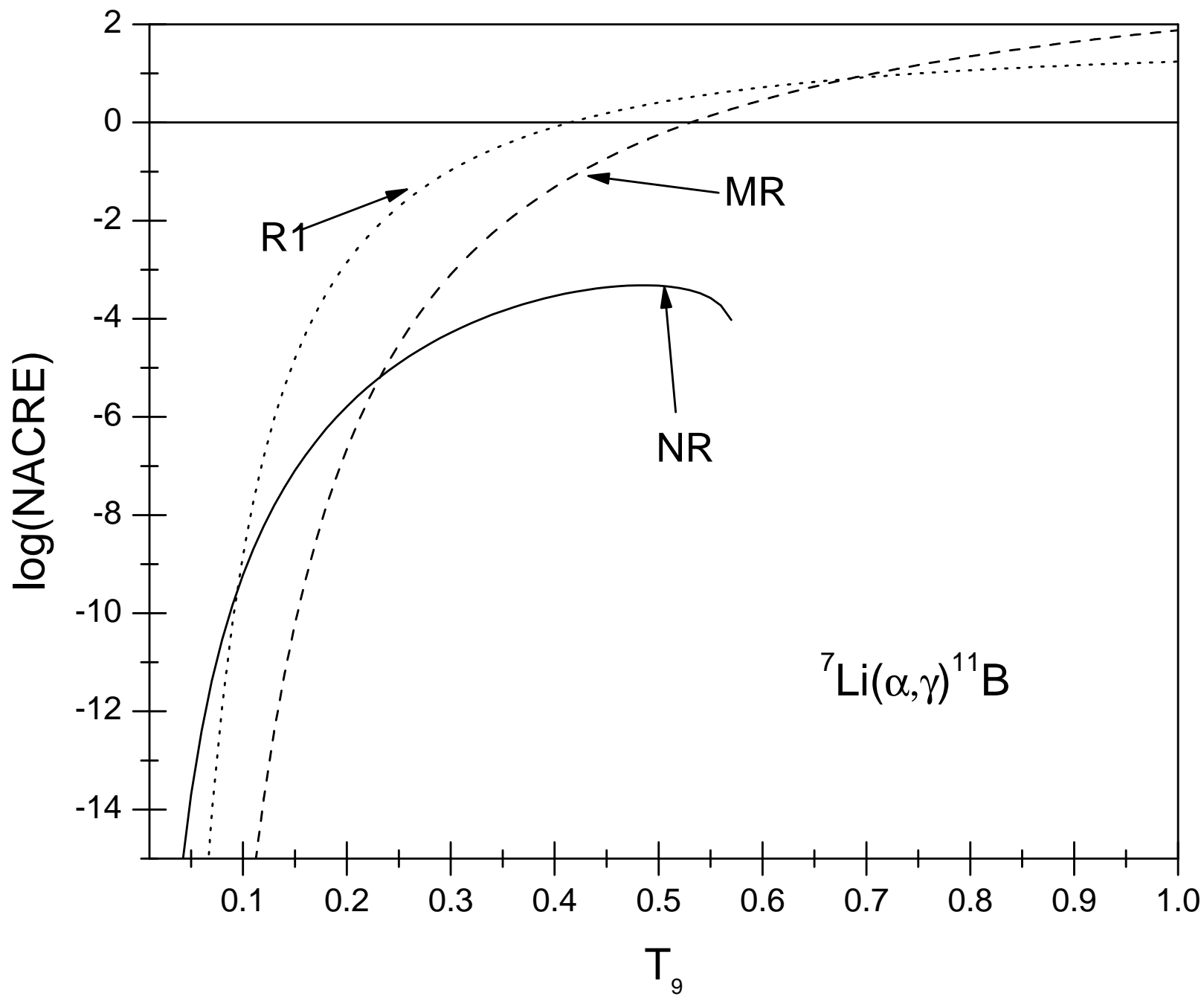
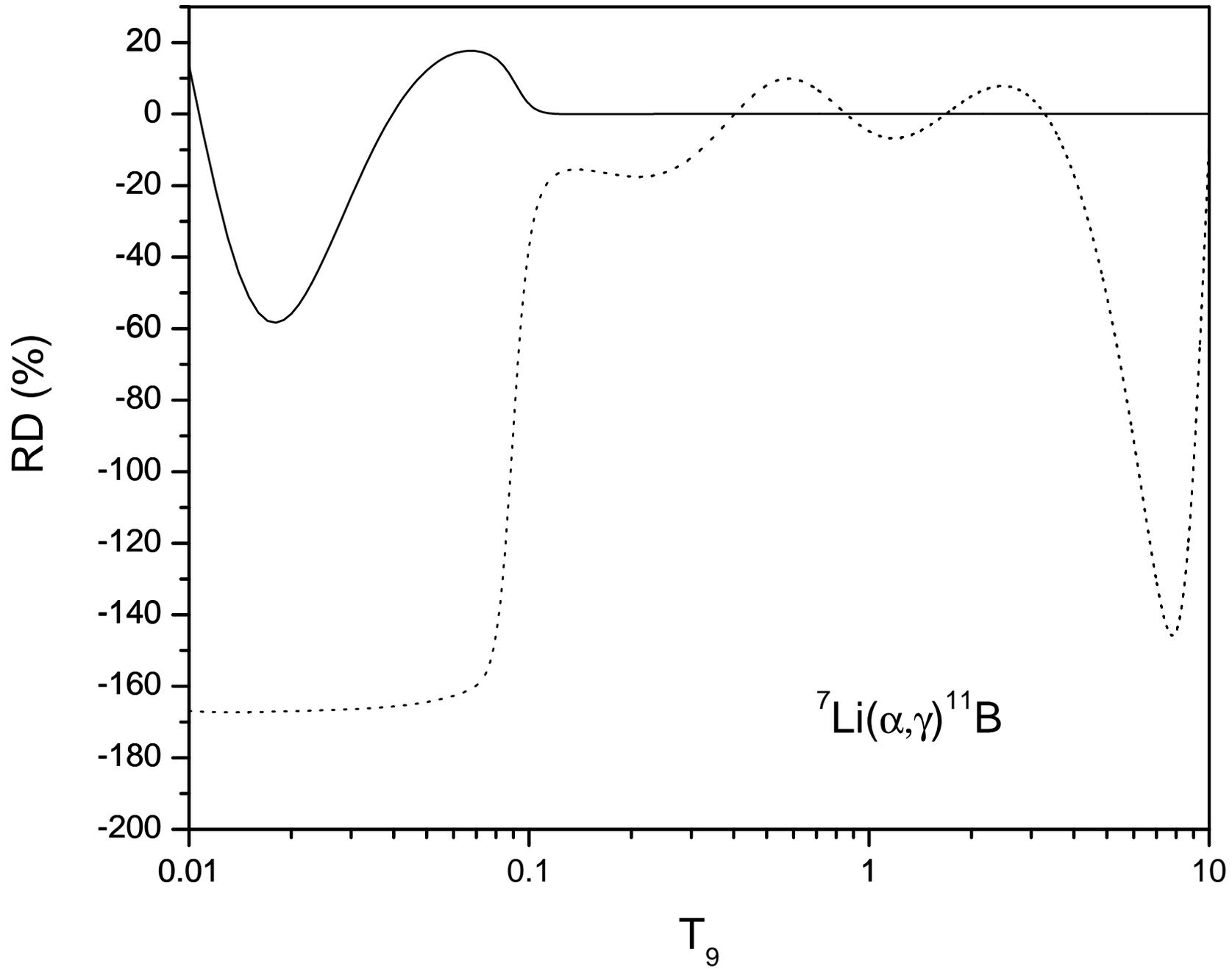
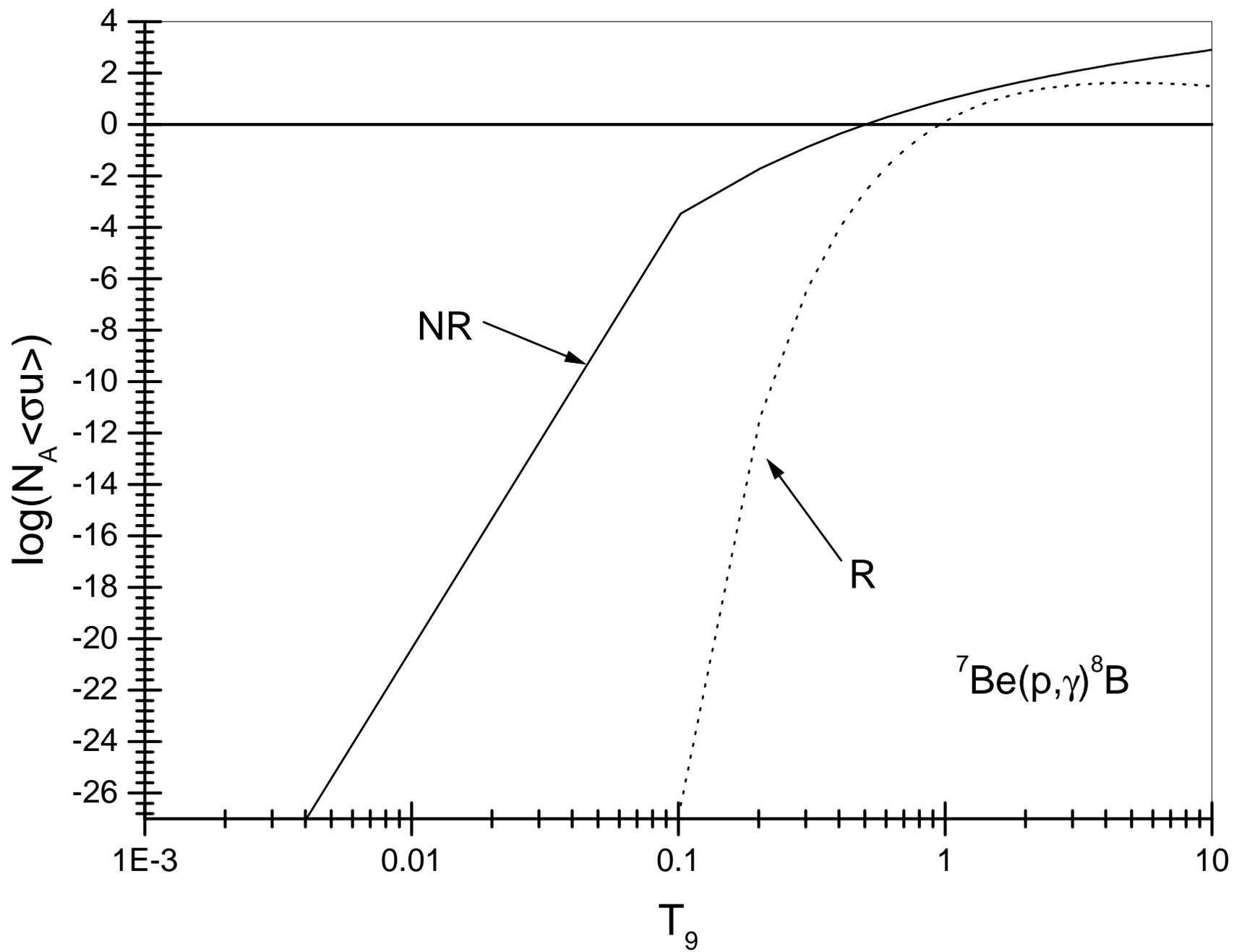
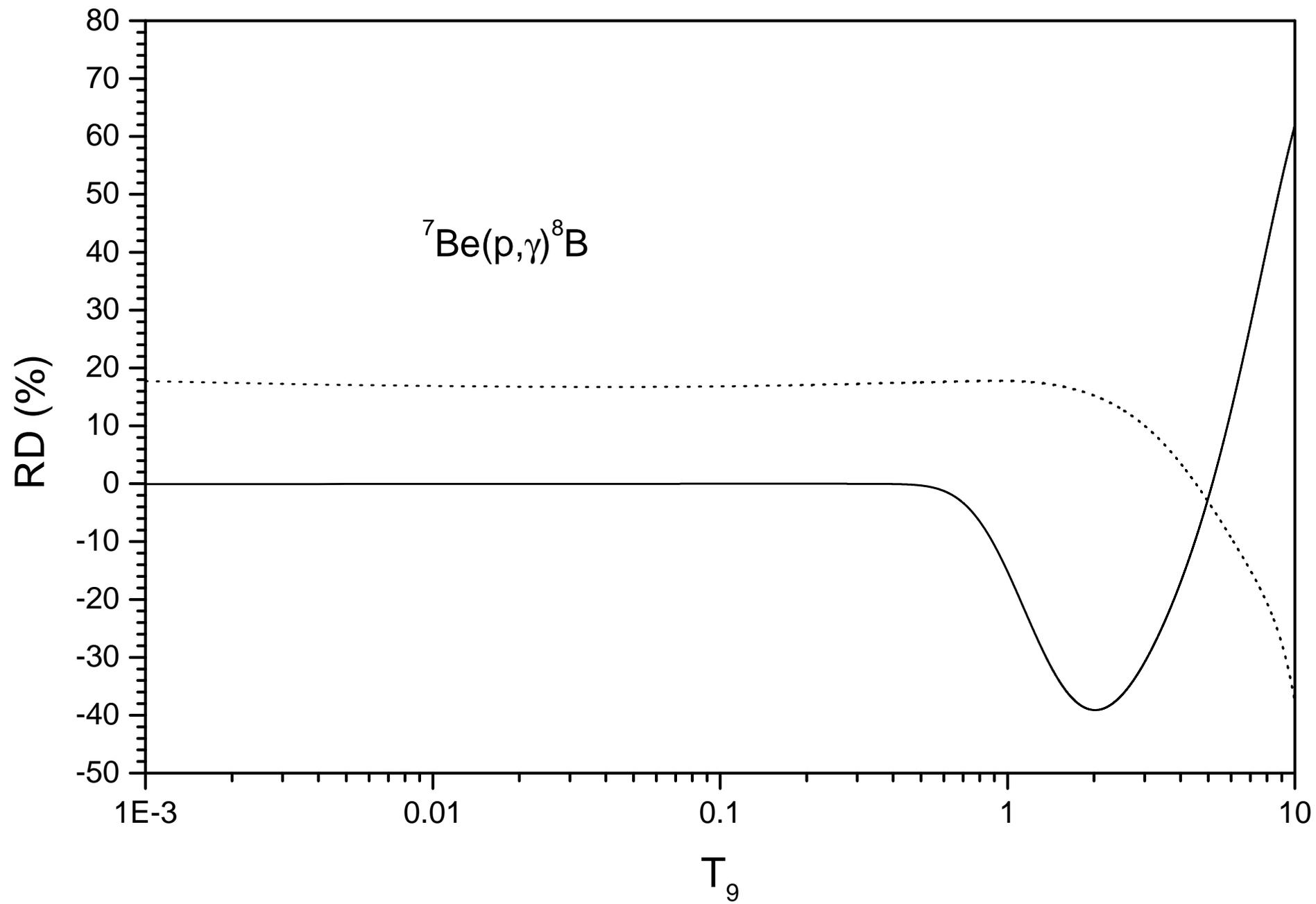
\end{document}